# The p53-MDM2 network: from oscillations to apoptosis

Indrani Bose and Bhaswar Ghosh

*Department of Physics, Bose Institute, 93/1, APC Road, Kolkata 700 009, India*

*\*Corresponding author (Fax, 91-33-2350 6790; Email, indbose@yahoo.co.in)*

The p53 protein is well-known for its tumour suppressor function. The p53-MDM2 negative feedback loop constitutes the core module of a network of regulatory interactions activated under cellular stress. In normal cells, the level of p53 proteins is kept low by MDM2, i.e. MDM2 negatively regulates the activity of p53. In the case of DNA damage, the p53-mediated pathways are activated leading to cell cycle arrest and repair of the DNA. If repair is not possible due to excessive damage, the p53-mediated apoptotic pathway is activated bringing about cell death. In this paper, we give an overview of our studies on the p53-MDM2 module and the associated pathways from a systems biology perspective. We discuss a number of key predictions, related to some specific aspects of cell cycle arrest and cell death, which could be tested in experiments.

## 1. Introduction

The p53 tumour suppressor protein is one of the most celebrated proteins because of its status as "guardian of the genome" (Vogelstein *et al* 2000). It constitutes the central node in a network of molecular interactions regulating the cellular response to stresses like DNA damage and oncogene activation. These stresses promote tumour formation, often culminating in cancer. The chief role of p53 is to guard cells against malignant transformation. When the cellular DNA is damaged by ionizing radiation or chemical agents, the appropriate p53-mediated pathways are activated. This results in the arrest of the cell division cycle which prevents the proliferation of cells containing damaged DNA (tumour formation). In the next step, the biochemical processes involved in DNA repair are initiated. Once this task is completed successfully, the cell resumes its progression so that cell division can take place. If repair is not possible due to excessive damage, the p53-mediated apoptotic pathway becomes functional, leading to apoptosis, i.e. programmed cell death.

The p53 activity is kept low in normal cells so that the cell cycle is not disrupted or the cell does not undergo untimely death. This is achieved through the operation of a negative feedback loop consisting of the p53 and MDM2 genes (Ma *et al* 2005). The p53 protein functions as a transcription factor (TF) and regulates the expression of several target genes. This is achieved through the binding of the p53 at the regulatory region of the target DNA. The MDM2 gene is one of the target genes the transcription of which is activated by the p53 proteins. The MDM2 proteins, however, inhibit the p53 activity, thus giving rise to a negative feedback loop. The ability of MDM2 to keep p53 in check is essential for normal cell function. The repression operates via three mechanisms. Firstly, MDM2 binds p53 at its DNA binding domain so that the latter cannot function as a TF. Secondly, MDM2 on binding p53 labels it for degradation and lastly, MDM2 is responsible for the export of p53 from the nucleus to the cytoplasm abrogating, in the process , its transcriptional activity. In mammalian cells, on DNA damage, a protein called ATM kinase is activated which phosphorylates the p53 protein at a specific site preventing the binding of the MDM2 protein to the p53 (Ghosh and Bose 2005). In the absence of MDM2 mediated degradation of the p53, the protein stabilizes at a higher level, i.e. is in the 'active' state. The vital role of the p53 protein in maintaining cellular integrity is underscored by the fact that the p53 gene is mutated in about 50% of human cancers. Systems biology approaches based on mathematical modelling provide significant insight on the operational aspects of the p53 mediated pathways. In this paper, we describe the mathematical models developed by us to study the p53-mediated cell cycle arrest and apoptosis. We discuss briefly the major results obtained and point out their experimental relevance.

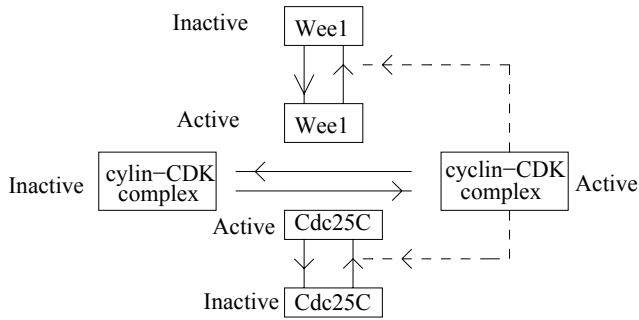

**Figure 1.** Molecular interactions of the cyclin-CDK complex with Wee1 and Cdc25C.

## 2. Cell cycle arrest

The cell cycle is a sequence of orderly events during which a growing cell duplicates all its components so that after cell division each daughter cell has the same cellular constituents as the parent cell (Alberts *et al* 2002). In an eukaryotic organism, the cell cycle progresses through four distinct phases: $G_1$, $S$, $G_2$ and $M$ (mitotic) phases. The division of the parent cell into two daughter cells occurs in the $M$ phase. The eukaryotic cell cycle operates through the sequential activation and deactivation of cyclin dependent protein kinases (CDKs). Protein kinases are enzymes which regulate the structure and/or the activity of target proteins by the transfer of phosphate molecules (phosphorylation). Similarly, protein phosphatases regulate biochemical activity by removing phosphate molecules from target molecules (dephosphorylation). CDK activation can occur only after a regulatory protein cyclin binds CDK and the complex is phosphorylated by CDK-activating kinases (CAKs). Even if these conditions are met, the CDK may be inactivated by inhibitory phosphorylation carried out by certain protein kinases. Protein phosphatases remove the inhibitory phosphate molecules so that the activity of the cyclin-CDK complex is triggered. A cycle of synthesis and degradation of cyclins in each cell cycle controls the periodic assembly and activation of the cyclin-CDK complex. In higher eukaryotic organisms, different cyclin-CDK complexes initiate different cell cycle events. The cell control system, however, operates on similar principles which may thus be assumed to be universal.

In the case of DNA damage by radiation or chemicals, the cell cycle is arrested at the damage checkpoints in the $G_1$ and $G_2$ phases. We have developed a mathematical model of the DNA damage checkpoint in the $G_2$ phase of frog egg and mammalian cell cycles (Ghosh and Bose 2005). The activation of the checkpoint arrests the cell cycle transition from the $G_2$ to the $M$ phase. The active cyclin-CDK complex phoshorylates key intracellular proteins which in turn initiate or control important cell cycle events. The kinase Wee1 suppresses the activity of the cyclin-CDK complex through inhibitory phosphorylation. The active cyclin CDK complex, on the other hand, inactivates its antagonist, the kinase Wee1 (figure 1). One thus has an effective positive feedback loop involving Wee1 and the cyclin-CDK complex. The phosphatase Cdc25C removes the inhibitory phosphate group from the inactive cyclin-CDK complex converting it into the active form. The active cyclin-CDK complex activates its friend Cdc25C, giving rise to another positive feedback loop. The combination of positive feedback loops and nonlinearity in the dynamics of the system give rise to bistability in a range of parameter values (Novak and Tyson 1993, 2003; Tyson *et al* 2001). The two stable steady states correspond to the $G_2$ (low activity of the cyclin-CDK complex) and the $M$ (high activity of the cyclin-CDK complex) phases, respectively. The transition from the lower to the upper state occurs when the cyclin threshold crosses a critical value. Since the amount of cyclin is correlated with the cell size or cell mass/DNA, the latter quantity can be treated as the parameter the changing of which triggers the $G_2/M$ transition. Experimental evidence for hysteresis has recently been obtained in a frog egg extract confirming earlier theoretical predictions (Sha *et al* 2003).

The cell cycle is an example of a dynamical system in which events unfold as a function of time. The time evolution of the system is described in terms of a set of differential equations. Each different equation is a rate equation which has the form

$$\frac{dx}{dt} = rate\,of\,(gain - loss), \qquad (1)$$

where $x$ is the amount of one type of biomolecule, say, active Cdc25C. The "gain" ("loss") terms describe processes which increase (decrease) the amount. In the case of active Cdc25C, $x$ is increased when active cyclin-CDK complex phosphorylates inactive Cdc25C into the active form. The amount $x$ is decreased due to the conversion of the active Cdc25C into the inactive form (figure 1). In the case of the $G_2/M$ transition, one writes down a differential equation for each of the key biomolecules namely, active/inactive cyclin-CDK complex, Wee1 and Cdc25C. The mathematical expressions for the "gain" and "loss" terms on the RHS (right hand side) of eq. (1) are derived on the basis of standard chemical kinetic schemes (law of mass action, Michaelis-Menten kinetics etc.). The reaction scheme shown in figure 1 leads to the following differential equations.

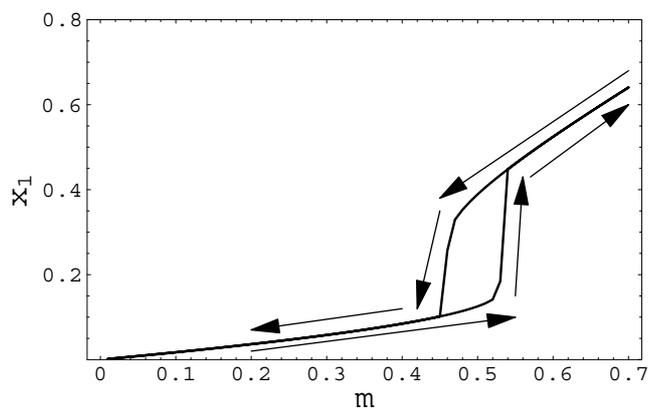

**Figure 2.** Hysteresis loop for the $G_2/M$ transistion; $x_1$ is the steady state concentration of the active cyclin-CDK complex and $m$ the cellular mass/DNA.

$$\frac{dx_1}{dt} = j_c m - k_1 y x_1 + k z x_2 - \gamma_{c1} x_1, \qquad (2)$$

$$\frac{dx_2}{dt} = k_1 y x_1 - k z x_2 - \gamma_{c2} x_2, \qquad (3)$$

$$\frac{dy}{dt} = k_2 \frac{1-y}{j_3 + 1 - y} - k'_2 x_1 \frac{y}{j_4 + y}, \qquad (4)$$

$$\frac{dz}{dt} = k_3 x_1 \frac{z_1}{j_5 + z_1} - k'_3 \frac{z}{j_6 + z} - \beta_1 z, \qquad (5)$$

$$\frac{dz_1}{dt} = \alpha - k_3 x_1 \frac{z_1}{j_5 + z_1} + k'_3 \frac{z}{j_6 + z} - \beta_1 z_1, \qquad (6)$$

where $x_1$ ($x_2$) denotes the concentration of the active (inactive) cyclin-CDK complex, $y$ is the concentration of the active Wee1 and $z$ ($z_1$) denotes the concentration of the active (inactive) Cdc25C. The total concentration of Wee1 is normalized to one so that $1 - y$ represents the concentration of the inactive Wee1. The $j_i$'s ($i = 1,....,6$) are the Michaelis-Menten constants. The first term in eq. (2) describes the synthesis of cyclin, the rate of which is proportional to the cell mass/DNA $m$. In the $G_2$ phase, cells are growing and larger cells, it is assumed, synthesize cyclin at a higher rate. The steady state solutions of eqs (2)–(6) $\left(\frac{dx}{dt} = 0\right)$ are obtained with the help of Mathematica. Figure 2 shows the result in the form of a hysteresis loop with $m$ playing the role of the bifurcation parameter. In the region of bistability, the two stable steady states correspond to the $G_2$ (lower branch) and $M$ phases (upper branch) respectively. When $m$ and correspondingly the cyclin concentration in the cyclin-CDK complex reaches a threshold value, the active complex inactivates Wee1 and activates Cdc25C in sufficient amounts. This triggers the autocatalytic conversion of the inactive cyclin-CDK complex into the active complex and a transition to the mitotic phase, with a higher concentration of the active cyclin-CDK complex, takes place. The results shown in figure 2 have been obtained for the following parameter values. The rate constants in appropriate units are $j_c=2$, $k = 1$, $k_1=1$, $\gamma_{c1}=1$, $\gamma_{c2}=1$, $k_2=0.4$, $k'_2= 2$, $k_3= 0.4$, $k'_3= 0.1$, $\alpha = 1$, $\beta_1 = 0.4$. The Michaelis-Menten constants are: $j_3 = 0.02$, $j_4 = 0.02$, $j_5 = 0.02$, $j_6 = 0.02$. Values of the different rate and Michaelis-Menten constants are consistent with those reported in earlier literature.

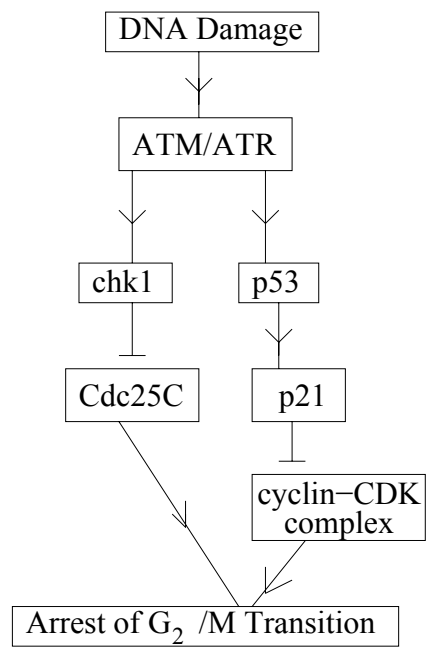

**Figure 3.** DNA damage response network leading to cell cycle arrest.

We now discuss the arrest of the cell cycle on DNA damage. The DNA damage response network is shown in figure 3. Recently, Lahav *et al* (2004) have performed single cell experiments on the dynamics of the network after DNA damage. The response of the network to the damage is found to be digital in the form of a discrete number of p53 and MDM2 protein pulses. As already mentioned, the p53-MDM2 network can be described in terms of a negative feedback loop. The simple motif can give rise to oscillations if the number of elements in the loop exceeds two or a time delay is included in the feedback process. A mathematical model of the p53-MDM2 negative feedback loop included an intermediary of unknown origin in order to obtain oscillations (Lev Bar-Or *et al* 2000). A later study proposed that the p53-MDM2 network includes both positive and negative feedback loops (Ciliberto *et al* 2005). The indirect inhibition of MDM2 by p53 in the positive feedback loop occurs via the interactions involving PTEN, PIP3 and Akt. The negative feedback loop also includes certain intermediate processes. It has been suggested that the intermediate processes in both the positive and negative feedback loops give rise to the time delay required for obtaining the MDM2 oscillations. We have proposed a simplified scheme in which oscillations are generated by considering the expression of the MDM2 gene to be a two step process, i.e. consisting of both transcription and translation. In the earlier modelling studies, only one step, namely, protein synthesis was taken into account. The two-step gene expression introduces a time delay in the p53-MDM2 network dynamics leading to oscillations (figure 4). The same result is obtained if time delay in the expression of both the p53 and MDM2 genes, via the two-step processes of transcription and translation, is considered. In our model, the pulses are obtained in the absence of the p53-MDM2 positive feedback loop provided the amount of DNA damage exceeds a threshold value. In the presence of the positive feedback loop, the magnitude of the threshold is lowered, i.e. the response occurs for a smaller amount of damage. The DNA damage response network in figure 3 shows that the arrest of the $G_2/M$ transition can occur via two pathways: either through the inhibition of Cdc25C activity by the chk1 kinase or via the inhibition of the cyclin-CDK complex activity by p21 proteins. The p53 pulses activate the transcription of the p21 gene. The differential equations describing the time evolution of the combined networks shown in figures 1 and 3 can be written down following standard procedure (Ghosh and Bose 2005). The solution of the equations yields interesting results. Figure 5 shows that the cell cycle is arrested (compare with figure 2) on DNA damage when the copy number of the p53 gene is two. The arrest implies a time delay in the transition from the $G_2$ to the M phase of the cell cycle. The delay increases, though not significantly, with increasing DNA damage. Eukaryotes are diploid organisms in which each gene has two identical copies. Knudson's well-known two hit model of tumourigenesis suggests that mutation in both copies (diploid organisms have two copies of each gene) of a tumour suppressor gene is essential for triggering tumour formation (Knudson 1971; Fodde and Smits 2002; Hohenstein 2004). Recent studies, however, show that the mutation of a single copy is sufficient in many cases for the loss of tumour suppression function of p53 protein. The gene dosage effect is called haploinsufficiency (HI) and has been verified experimentally (Venkatachalam *et al* 1998; Song *et al* 1999). Tumours are found to arise in mice with only one intact copy of the p53 gene contrary to Knudson's hypothesis. Our model calculations show that the $G_2/M$ transition is nor arrested when the copy number of the p53 gene is one. This results in a proliferation of cells containing damaged DNA which may result in cancer. The key theoretical predictions are, however, yet to be verified in an experiment which explicitly confirms the underlying mechanism based on bistability.

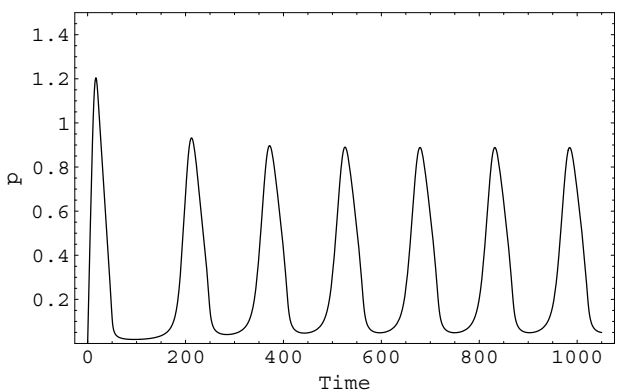

**Figure 4.** Pulses of p53 proteins generated on DNA damage; p is the total p53 concentration.

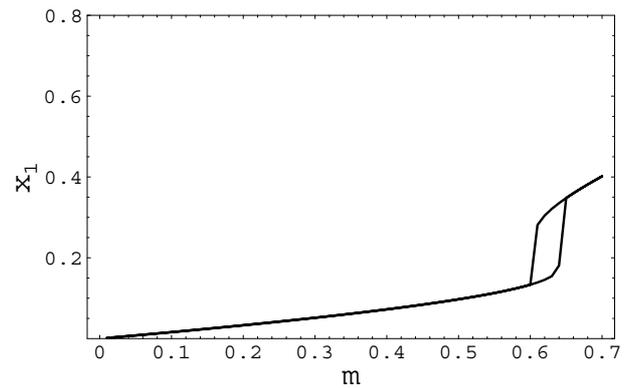

**Figure 5.** Cell cycle arrest on DNA damage; $x_1$ is the steady state concentration of the action cyclin-CDK complex and *m* the cellular mass/DNA.

### 3. Proliferation versus death

MDM2, the biological regulator of p53, is found to be overexpressed in certain types of cancer (Tovar *et al* 2006). The origin of increased gene expression lies in gene amplification or enhanced rates of transcription and translation. Since MDM2 is a negative regulator of p53 activity, the overproduction of MDM2 proteins gives rise to a greater risk in the development of malignancy. To counteract this, one strategy is to use small-molecule inhibitors to block the protein-protein interaction of MDM2 and p53. Vassilev and coworkers (Vassilev *et al* 2004; Tovar *et al* 2006) have developed a class of small molecules known as nutlins which occupy the p53 binding pocket in MDM2, thus preventing the binding of MDM2 to p53 and facilitating the activation of the p53 pathways in human cancer cell lines. The efficacy of the strategy has been demonstrated in experiments paving the way for the use of nutlins as an anticancer drug. Vassilev *et al* (2004) selected a panel of 10 cell lines, derived from different human tumours, for their experiments. In all the cell lines, the p53 genes are not mutated. In the osteosarcoma (SJSA-1) cell line, the MDM2 gene is amplified 25-fold whereas all the other cell lines have a single copy of the MDM2 gene. Exponentially growing cells in all the cancer cell lines were treated with nutlin-3 for several hours. The cell cycle progression was found to be arrested in all the cell lines. The apoptotic response, however, was significant (~ 80%) only in SJSA-1, i.e. in the case of MDM2 overexpression. The response was low in the other cell lines, e.g. <10% in HCT116 (colorectal cancer). A high apoptotic response was also obtained in a second osteosarcoma cell line (MHM) with an ≈ 10-fold amplification of the MDM2 gene. We have developed a mathematical model to explore the possible origin of the considerable difference in the apoptotic response of cancer cell lines, with and without MDM2 overexpression, when subjected to nutlin-3 treatment. Some aspects of the model are similar to those of earlier mathematical models of the apoptotic pathways (Fussenegger *et al* 2000; Bagci *et al* 2006). In the following, we give a brief description of the major results obtained. A fuller account of the study will be published elsewhere (Ghosh B and Bose I, unpublished results).

Figure 6a shows a schematic diagram of the p53-mediated apoptotic pathway which is activated when the p53-MDM2 interaction is inhibited by nutlin molecules. The p53 proteins upregulate the synthesis of the pro-apoptotic Bax proteins and downregulate that of the anti-apoptotic Bcl-2 proteins (Fussenegger *et al* 2000; Hengartner 2000; Bagci *et al* 2006; Ghosh B and Bose I, unpublished results). The Bcl-2 binds the Bax protein to form a heterodimer and consequently suppress the pro-apoptotic activity of Bax. The Bcl-2 further binds the apoptotic protease-activity factor 1 (Apaf-1) and thereby excludes the binding of procaspase-9 to Apaf-1. The Bax and other pro-apoptotic proteins help in the release of cytochrome c proteins from the mitochondria. Apaf-1 binds cytochrome c to form the apoptosome complex which in turn binds procaspase-9. A family of proteases termed caspases play a central role in bringing about apoptosis. The blocking of caspase activity can slow down or prevent apoptosis. Caspases selectively cleave a set of specific target proteins at one or a few positions, always after an aspartate residue. The cleavage results in non-functional target proteins. The binding of the procaspase-9 to the apoptosome leads to initiator caspase-9 activation through cleavage of the bound procaspase-9. The initiator caspases activate the executioner or effector caspases. This is followed by the cleavage of procaspase-3 to yield the active effector molecule, caspase-3. Executioner caspases like caspase-3 cleave key cellular proteins and dismantle the cell resulting in apoptosis. The end comes provided the executioner caspases such as caspase-3 are activated to sufficiently high levels. The level of activation depends on the competition between the pro- and anti-apoptotic agents/processes constituting the apoptotic pathway shown in figure 6a. We have retained only the key components in the p53-mediated apoptotic network which in reality is quite complex. The apoptotic network further includes two important processes, namely, the cleavage of the anti-apoptotic Bcl-2 by caspase-3 and the inhibition of the processing and activation of procaspase-3 by p21. These two processes provide direct links between the events upstream and downstream of cytochrome c release. There is now increasing evidence that p21 has a dual role in the p53-MDM2 network of a number of systems (Gartel and Tyner 2000; Dotto 2000). The p21 protein levels are activated under cellular stress due to the activation of p53 proteins. The protein helps in the arrest of the cell cycle and may also have the dual function of protecting cells against apoptosis. Figure 6b shows some additional biochemical processes incorporated in the p53-MDM2 network of our model to study apoptosis. One of these relates to the observation that MDM2 is a negative regulator of p21 independent of p53. The physical interaction between p21 and MDM2, leading to the binding of the proteins both *in vitro* and *in vivo*, has been demonstrated in experiments (Zhang *et al* 2004). MDM2 promotes p21 degradation by facilitating the binding of a proteasomal subunit. Thus, MDM2 overexpression gives rise to lower p21 protein levels. Experiments on the HCT116 human colorectal cancer cells (one of the cell lines in Vassilev *et al*'s experiment) show that the p53 levels are enhanced with abolition of the p21 gene expression (Javelaud and Besançon 2000). In fact, the Bax/Bcl-2 ratio has been found to be 4-fold higher in the absence of p21 proteins. A higher Bax/Bcl-2 ratio tilts the balance in the favour of apoptosis. Disruption of p21 activity leads to enhanced $p14^{Arf}$ expression. The p14 proteins are antagonistic to the MDM2 activity so that p53 proteins can accumulate to higher levels.

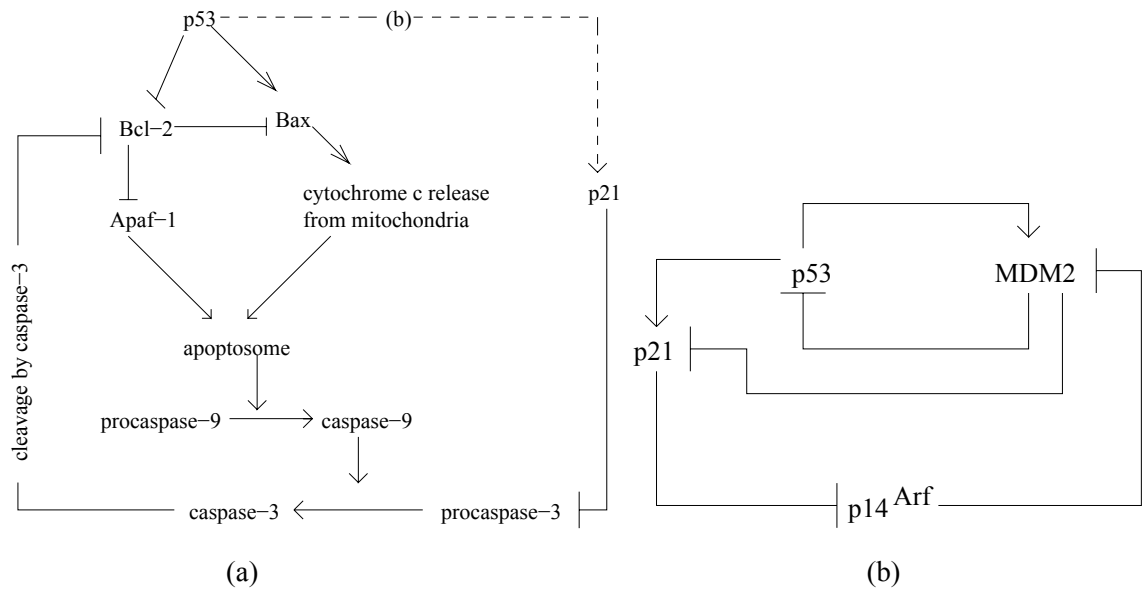

**Figure 6.** (**a**) The p53-mediated apoptotic pathway. (**b**) the subnetwork of regulatory interactions between p53, MDM2 and p21. The dotted link in (**a**) includes the subnetwork shown in (**b**).

In the mathematical model describing the p53-mediated apoptotic pathway (figure 6), the time evolution of the system can again be represented by a set of coupled differential equations (Ghosh B and Bose I, unpublished results). The steady state solutions of the equations for different parameter values are obtained numerically. Figure 7 shows a generalised plot of the steady state amount of caspase-3 versus the amounts of p53 and p21. The latter two concentrations are treated as independent variables. In the steady state of the p53-MDM2 network shown in figure 6b, the concentrations of the p53 and p21 proteins can vary over wide ranges depending on the magnitudes of the different parameters of the network. The latter include various rate and binding constants, the basal expression levels and the gene copy number of, say, the MDM2 gene. These quantities have values which are often cell and tissue-specific so that diverse combinations of the steady state concentrations of p53 and p21 are possible. The dotted link between p53 and p21 in figure 6a includes the subnetwork shown in figure 6b. The plot in figure 7 is obtained with the p53 and p21 proteins serving as independent inputs to the apoptotic pathway in figure 6a. The two protein amounts are given by the respective concentrations in the steady state of the subnetwork shown in figure 6b. The apoptotic pathway has a comparatively slower dynamics so that the p53-MDM2 subnetwork reaches the steady state earlier. The goal is to study caspase-3 activation as a function of the p53 and p21 protein concentrations. The plot in figure 7 indicates a threshold-type activation of the caspase-3 level. We now focus attention on the two cancer cell lines, SJSA-1 (MDM2 overexpressed) and HCT116 (MDM2 not overexpressed). In both the cases, the steady state concentrations of the p53 and p21 proteins are determined from the set of differential equations governing the time evolution of the subnetwork shown in figure 6b. In the case of SJSA-1 cells, the level of p21 is found to be low since the degradation of p21 is enhanced due to an overexpression of the MDM2 proteins (Ghosh B and Bose I, unpublished results). The negative influence of p21 on the p53 and caspase-3 levels is less prominent in this case. From the generalized plot, one finds that the caspase-3 level, for the specific steady state amounts of p53 and p21, is sufficiently activated so that apoptosis is possible. In the case of HCT116, the p21 level is high due to the lesser amount of degradation by MDM2 so that apoptosis is not favoured. We have also modeled the activation of p53 on treatment with nutlin and find that the cell cycle is arrested in both the SJSA-1 and HCT116 cells (Ghosh B and Bose I, unpublished results).

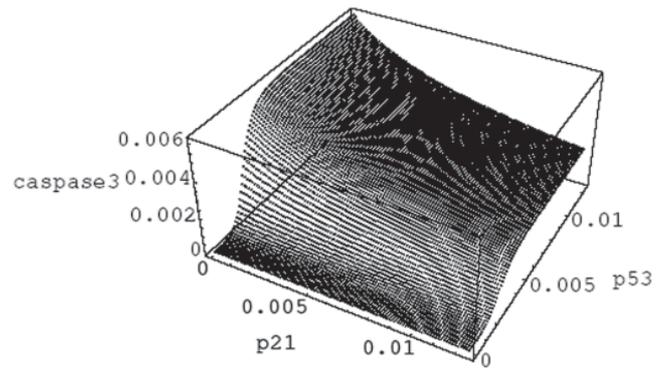

**Figure 7.** Steady state concentration of caspase-3 versus p53 and p221 amounts. The letter two proteins serve as independent inputs to the apoptotic pathway shown in figure 6a.

In summary, we have described the salient features of our modelling studies on the p53-MDM2 network from a systems biology perspective. We have shown that the cell cycle is not arrested on DNA damage when the copy number of the p53 tumour suppressor gene is reduced from two to one. This prediction could be tested in an appropriately designed experiment. We have further shown that the experimentally observed p53 and MDM2 oscillations can be explained on the basis of time delay caused by the two-step nature (transcription and translation) of gene expression. The importance of time delay in the functional response of the p53-MDM2 network has been pointed out in some related studies (Tiana *et al* 2002; Wagner *et al* 2005).

There is a marked difference in the apoptotic response of cancer cells with normal MDM2 expression and MDM2 overexpression when treated with nutlin, an inhibitor of the p53-MDM2 interaction. A possible explanation of the observed difference has been put forward in the framework of a mathematical model. The model shows a threshold-type activation of the caspase-3 level as a function of p53 and p21 levels. Low levels of caspase-3 cannot bring about cell death. The amount of p21, the transcription of which is activated by p53, appears to be a crucial factor in determining the cell fate. The transcription of the p21 gene is activated by p53 proteins whereas the MDM2 proteins promote the degradation of the p21. The p21 has a negative effect on both the p53 and caspase-3 levels. If the p53 level is low, the ratio of the amounts of Bax and Bcl-2 proteins is not sufficiently high to favour the release of cytochrome c from the mitochondria. The activation of the caspase-3 level is hence minimal. Some of the ideas put forward can be tested experimentally by varying the p21 levels in cancer cells like SJSA-1 and HCT116 during nutlin treatment.

**Acknowledgment**

BG is supported by the Council of Scientific and Industrial Research, New Delhi under section number 9/15 (282)/2003-EMR-1.